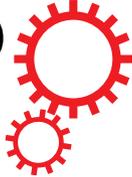

# Current driven spin–orbit torque oscillator: ferromagnetic and antiferromagnetic coupling

Øyvind Johansen & Jacob Linder

We consider theoretically the impact of Rashba spin–orbit coupling on spin torque oscillators (STOs) in synthetic ferromagnets and antiferromagnets that have either a bulk multilayer or a thin film structure. The synthetic magnets consist of a fixed polarizing layer and two free magnetic layers that interact through the Ruderman-Kittel-Kasuya-Yosida interaction. We determine analytically which collinear states along the easy axis that are stable, and establish numerically the phase diagram for when the system is in the STO mode and when collinear configurations are stable, respectively. It is found that the Rashba spin–orbit coupling can induce anti-damping in the vicinity of the collinear states, which assists the spin transfer torque in generating self-sustained oscillations, and that it can substantially increase the STO part of the phase diagram. Moreover, we find that the STO phase can extend deep into the antiferromagnetic regime in the presence of spin–orbit torques.

Twenty years ago, it was theoretically proposed by Slonczewski[1] and Berger[2] that there could be exerted a torque on the magnetization in multilayer systems by passing a spin polarized current through the magnetic layers. This was coined the spin-transfer torque (STT) as the spin of the spin polarized current was transferred to the magnetic layer. This effect was observed experimentally[3,4] a few years after the publications by Slonczewski and Berger, and spiked a lot of interest in the field as the magnetization could now be manipulated by electrical means, which is often advantageous practically compared to manipulation by magnetic fields. This torque could be used to switch the magnetization direction in one of the magnetic layers above some critical current[4–6], which is of interest for writing techniques in memory technologies such as MRAM[7,8] and racetrack memories[9]. The spin-transfer torque was also shown to induce a precession in the free magnetic layers[10–12], which is now known as a spin torque oscillator (STO). The spin torque oscillator takes in a dc spin polarized current, and due to the precession in the free magnetic layers that causes an oscillation in the resistance through the giant magnetoresistance effect[13,14], and the result is an ac current passing out of the multilayers. These alternating currents can have a wide range of frequencies, spanning the range of 100s of MHz to 100s of GHz[8,15,16], and these frequencies are tunable by the magnitude of the applied current. Spin torque oscillators can exist in both antiferromagnetically[17] and feromagnetically[18] coupled magnetic layers, although it was noted in ref. 18 that they were unable to reproduce the antiferromagnetic STO phase predicted in ref. 17. Antiferromagnetic nano-oscillators were also recently considered in ref. 19.

Another type of torque that has gained interest in magnetization dynamics in more recent years is the torque resulting from Rashba spin–orbit coupling (RSOC)[20]. This type of spin–orbit coupling occurs in materials with broken inversion symmetry, such as at the interface between two materials[21]. This inversion asymmetry causes an in-plane current flowing parallel to the interface to experience a magnetic field perpendicular to both the direction of the current and inversion asymmetry[22]. Rashba spin–orbit coupling has been shown to introduce interesting effects in many different areas of physics[23], and is of particular interest due to the fact that the strength of the interaction can be tuned by gate voltages[24,25]. RSOC can, like STT, be utilized in magnetization switching, although RSOC is not the mechanism solely responsible for it[26]. Several works have considered the influence of spin–orbit torques on magnetic domain wall motion[27–37]. It has also been observed experimentally that the spin–orbit torque from RSOC can contribute to self-oscillations in STOs[38].

In this article, we show that RSOC can be used in metallic multilayer systems to substantially increase the size of the STO phase. Moreover, we discover an STO phase for two compensated antiferromagnetically coupled magnetic layers, which is a new result compared to *e.g.* Zhou *et al.* who could only find an STO phase

Department of Physics, NTNU, Norwegian University of Science and Technology, N-7491 Trondheim, Norway. Correspondence and requests for materials should be addressed to J.L. (email: jacob.linder@ntnu.no)

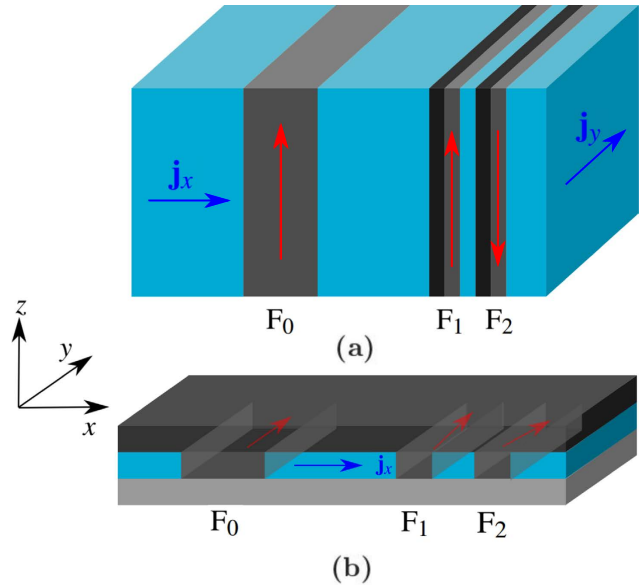

**Figure 1.** Illustrations of (**a**) the bulk geometry and (**b**) the thin film geometry. In both geometries, a fixed magnetic layer $F_0$ is separated from two free magnetic layers $F_1$ and $F_2$ by a non-magnetic metallic material shown here in blue. A material illustrated in black, which is neighboring to $F_1$ and $F_2$ in (**a**) and the top film in (**b**), is present to get a strong Rashba spin–orbit coupling at the interface of the free ferromagnetic layers. A suitable choice for this material could be a heavy normal metal such as Au or Pt. In (**b**) the bottom film is a substrate which we assume does not induce any measurable interfacial spin–orbit coupling effects. To induce dynamics a current is applied in the $x$-direction which causes $\mathbf{m}_1$ and $\mathbf{m}_2$ to experience spin-transfer torques. In (**a**) a current is also applied in the $y$-direction to create significant RSOC effects on $\mathbf{m}_1$ and $\mathbf{m}_2$ due to the symmetry breaking in the $x$-direction. In the film geometry this is caused by the current in the $x$-direction as the symmetry breaking is in the $z$-direction. The free magnetic layers also interact through the RKKY interaction, while the distance to the fixed magnetic layer is chosen such that an RKKY interaction with this layer can be neglected. In (**a**) the material has an easy axis in the $z$-direction, while in (**b**) the material has an easy axis in the $y$-direction.

for a ferromagnetic coupling in ref. 18, and Klein *et al.* could only find an STO phase for uncompensated antiferromagnetically coupled magnetic layers in ref. 17. We begin by setting up our model by utilizing the Landau-Liftshitz-Gilbert-Slonczewski (LLGS) equation, and then proceed by performing a Fourier transform of this equation to find when collinear states along the easy axis are stable. By comparing the new terms introduced by having RSOC present, we find that the spin–orbit torques effectively can be described by a modification of the Gilbert damping $\alpha$, to the extent where we can get an anti-damping term in the LLGS equation. To establish when we have an STO phase we solve the full LLGS equation numerically for different sets of experimentally relevant parameters, considering two possible geometries, and use the solutions to classify the phases. Lastly, we analyze the frequency spectrum of the STO phases by performing a Fourier transform of the solutions along different lines in the phase diagrams.

## Theory

We will consider two different geometries where spin–orbit torques strongly influence the STO phase, and which display different behaviors. These geometries will henceforth be called the bulk and thin film geometries, and are illustrated in Fig. 1. Both geometries consist of a polarizing layer $F_0$ and two free magnetic layers $F_1$ and $F_2$, all separated by a non-magnetic metal in order to reduce the exchange coupling and prevent magnetic locking. The main differences between the geometries is the direction of which the inversion symmetry is broken, and the current that is required to induce the spin–orbit coupling. In the bulk geometry the current in the $y$-direction induces spin–orbit torques (SOT) from RSOC, while the current in the $x$-direction induces the STT. Achieving current injection in two directions could potentially be challenging to achieve experimentally, but the possibility of separating the effects of STT and SOT by control of different current directions is an appealing concept that we here put forward in order to stimulate to experimental activity on such a setup. No such difficulty occurs in the thin film setup where there is only current injection in one direction. In the thin film geometry both STT and RSOC are caused by the same current in the $x$-direction, making it impossible so separate the effects from one another. In contrast, this is possible in the bulk geometry. The thin film geometry may also prove to be experimentally challenging, as it is fair to acknowledge the difficulties of electrically tuning the Rashba parameter in a well-defined manner in this geometry.

The model for the dynamics of the magnetizations in the two free magnetic layers that we will study in this article is a variation of two coupled Landau–Lifshitz–Gilbert–Slonczewski (LLGS) equations. This equation is given by

$$\partial_t \mathbf{m}_i = -\gamma \mathbf{m}_i \times (\mathbf{H}_i^{\text{eff}} + \mathbf{H}_i^{\text{R}} - \beta \mathbf{m}_i \times \mathbf{H}_i^{\text{R}}) + \alpha \mathbf{m}_i \times \partial_t \mathbf{m}_i + \mathbf{T}_i^{\text{STT}}. \tag{1}$$

Here the Rashba field $\mathbf{H}_i^R$ is given by[27]

$$\mathbf{H}_i^R = \frac{1}{1+\beta^2} \frac{\alpha_R m_e P_i^{(x/y)} j_{(x/y)}}{eM_s \hbar \mu_0} \hat{\mathbf{n}} \times \hat{\mathbf{j}}, \qquad (2)$$

with $\alpha_R$ being the Rashba parameter, $P_i^{(x/y)}$ the polarization of the current $j_{(x/y)}$ passing through $F_i$, and $\hat{\mathbf{n}}$ the direction of inversion asymmetry. The parameter $\beta$ is the non-adiabatic damping parameter of the itinerant electrons that describes the ratio of the inverse exchange interaction parameter $1/J_{ex}$ and the spin-flip relaxation time. Lastly, the spin-transfer torques $\mathbf{T}_i^{STT}$ are acting on $\mathbf{m}_i$ are

$$\mathbf{T}_1^{STT} = -\frac{\gamma \hbar j_x}{2eM_s \mu_0 d} \mathbf{m}_1 \times (P_0 \mathbf{m}_0 - P_1 \mathbf{m}_2) \times \mathbf{m}_1, \qquad (3)$$

$$\mathbf{T}_2^{STT} = -\frac{\gamma \hbar j_x}{2eM_s \mu_0 d} P_1 \mathbf{m}_2 \times \mathbf{m}_1 \times \mathbf{m}_2, \qquad (4)$$

with $d$ being the thickness of the regions $F_1$ and $F_2$ in the $x$-direction, and $P_0$ and $P_1$ being the polarization of the current in the non-magnetic material in $F_0/N/F_1$ and $F_1/N/F_2$ respectively. Due to the symmetry of the geometry we can neglect the non-adiabatic spin transfer torque[39]. For the effective field $\mathbf{H}_{eff}$, we consider contributions from the RKKY interaction and magnetic anisotropy. The effective field then becomes

$$\mathbf{H}_i^{eff} = \frac{2K}{\mu_0 M_s} (\mathbf{m}_i \cdot \hat{\mathbf{n}}_k) \hat{\mathbf{n}}_k + \frac{J}{\mu_0 M_s d} \mathbf{m}_{\bar{i}}. \qquad (5)$$

$K$ is the anisotropy strength, $\hat{\mathbf{n}}_k$ a direction along the easy axis ($\hat{\mathbf{n}}_k = \hat{\mathbf{z}}$ for the bulk geometry, and $\hat{\mathbf{n}}_k = \hat{\mathbf{y}}$ for the thin film geometry), $J$ is the strength of the RKKY interaction, and the index $\bar{i}$ denotes the index of the free magnetization that is not $\mathbf{m}_i$ ($\bar{i} = 3 - i$).

We now follow the procedure by Zhou *et al.*[18] and consider when the collinear states of $\mathbf{m}_1$ and $\mathbf{m}_2$ along the easy axis are stable or not. We start off with the ansatz that there is a slight perturbation $\mathbf{u}_i$ from the collinear state, such that $\mathbf{m}_i = \lambda_i \hat{\mathbf{n}}_k + \mathbf{u}_i$ ($\lambda_i = \pm 1$). Plugging this ansatz into (1) and performing a Fourier transform $\mathbf{u}_i(t) = \int \tilde{\mathbf{u}}_i(\omega) \exp(-i\omega t) d\omega/2\pi$, we get a result that is on the form

$$(\hat{A}\omega + \hat{V}) \begin{pmatrix} \tilde{\mathbf{u}}_1 \\ \tilde{\mathbf{u}}_2 \end{pmatrix} = 0. \qquad (6)$$

The matrices for the different geometries are given by

$$\hat{A}_{bulk} = \begin{pmatrix} 1 - i\alpha\lambda_1 & 0 \\ 0 & 1 - i\alpha\lambda_2 \end{pmatrix}, \qquad (7)$$

$$\hat{A}_{film} = \begin{pmatrix} 1 + i\alpha\lambda_1 & 0 \\ 0 & 1 + i\alpha\lambda_2 \end{pmatrix}, \qquad (8)$$

$$\begin{aligned}\hat{V}_{bulk} =\ & \omega_0 \begin{pmatrix} \lambda_1 & 0 \\ 0 & \lambda_2 \end{pmatrix} + \omega_J \begin{pmatrix} \lambda_2 & -\lambda_1 \\ -\lambda_2 & \lambda_1 \end{pmatrix} + i\omega_j^{(x)} \left[ P_0 \begin{pmatrix} -\lambda_1 & 0 \\ 0 & 0 \end{pmatrix} \right. \\ & \left. + P_1 \begin{pmatrix} \lambda_1 \lambda_2 & -1 \\ 1 & -\lambda_1 \lambda_2 \end{pmatrix} \right] + \frac{\omega_R^{(y)}}{1+\beta^2} \left[ \begin{pmatrix} P_1^{(y)} & 0 \\ 0 & P_2^{(y)} \end{pmatrix} \right. \\ & \left. - i\beta \begin{pmatrix} P_1^{(y)} \lambda_1 & 0 \\ 0 & P_2^{(y)} \lambda_2 \end{pmatrix} \right], \end{aligned} \qquad (9)$$

$$\begin{aligned}\hat{V}_{film} =\ & -\omega_0 \begin{pmatrix} \lambda_1 & 0 \\ 0 & \lambda_2 \end{pmatrix} - \omega_J \begin{pmatrix} \lambda_2 & -\lambda_1 \\ -\lambda_2 & \lambda_1 \end{pmatrix} + i\omega_j^{(x)} \left[ P_0 \begin{pmatrix} -\lambda_1 & 0 \\ 0 & 0 \end{pmatrix} \right. \\ & \left. + P_1 \begin{pmatrix} \lambda_1 \lambda_2 & -1 \\ 1 & -\lambda_1 \lambda_2 \end{pmatrix} \right] - \frac{\omega_R^{(x)}}{1+\beta^2} \left[ \begin{pmatrix} P_1^{(x)} & 0 \\ 0 & P_2^{(x)} \end{pmatrix} \right. \\ & \left. + i\beta \begin{pmatrix} P_1^{(x)} \lambda_1 & 0 \\ 0 & P_2^{(x)} \lambda_2 \end{pmatrix} \right]. \end{aligned} \qquad (10)$$

Here, we have defined the frequencies $\omega_0 = 2\gamma K/\mu_0 M_s$, $\omega_J = \gamma J/\mu_0 M_s d$, $\omega_j^{(x)} = \gamma \hbar j_x/2\mu_0 M_s d$ and $\omega_R^{(x/y)} = \gamma \alpha_R m_e j_{x/y}/\hbar \mu_0 e M_s$. In general there are also off-diagonal terms in the $\hat{A}$ matrices, which can be attributed to anisotropic damping terms such as spin pumping[40]. However, similarly to[18] we neglect these terms to reduce the amount of parameters in our model.

We now want to determine if any of the collinear states are stable when the frequencies and other constants are specified. This can be seen from the sign of the imaginary component of $\omega$; when the imaginary component is negative, $\exp(-i\omega t)$ is a decreasing function in time, while if the imaginary component is positive the function is exponentially increasing. Any small perturbation away from the collinear state is then unstable if $\Im(\omega) > 0$. The value of $\omega$ can be determined from (6), as it can be written as an eigenvalue equation where $\omega$ is an eigenvalue of the matrix $\hat{W} = -\hat{A}^{-1}\hat{V}$. From this one can then find for what choice of parameters none of the collinear states are stable. An STO phase will be localized within this region, but the entire region is not necessarily an STO phase. We here distinguish between the STO phase and a canted phase by considering the temporal evolution of the magnetoresistance, which is approximated by

$$R(t) = R_0 + \Delta R_1 \mathbf{m}_0 \cdot \mathbf{m}_1 + \Delta R_2 \mathbf{m}_1 \cdot \mathbf{m}_2. \quad (11)$$

If the magnetoresistance is oscillating in time, meaning at least $\mathbf{m}_0 \cdot \mathbf{m}_1$ or $\mathbf{m}_1 \cdot \mathbf{m}_2$ is oscillating, we have an STO phase. If $\mathbf{m}_0 \cdot \mathbf{m}_1$ and $\mathbf{m}_1 \cdot \mathbf{m}_2$ are both constant, we have a canted phase. Note that if $\mathbf{m}_1$ and $\mathbf{m}_2$ oscillate in-phase in a plane perpendicular to $\mathbf{m}_0$, such that $\mathbf{m}_0 \cdot \mathbf{m}_1$ and $\mathbf{m}_1 \cdot \mathbf{m}_2$ are constant, we still have a canted phase and not an STO phase even though there are oscillations in the individual magnetization components. To confirm our analytical results, we will also later establish fully numerically when the STO phase occurs.

We now want to determine the physical effect that the Rashba spin–orbit coupling introduces by comparing it with other known terms. If we make the simplification of an equal polarization in the current causing the spin–orbit coupling in the free layers ($P_0 = P_1 = P$, $P_1^{(y)} = P_2^{(y)} = P$, $P_1^{(x)} = P_2^{(x)} = P$), this can be done without much effort. Note that for the bulk case it is not necessary to have the same polarization for both $j_x$ and $j_y$, we could still achieve the same results with $P_1^{(y)} = P_2^{(y)} = P'$ by modifying $\alpha_R$ so that $\alpha_R P = \alpha'_R P'$. When performing this simplification in the polarization the matrix proportional to $\omega_R^{(x/y)}$ becomes very similar to the $\hat{A}$ matrix. The influence of RSOC is then to renormalize $\alpha$ and $\omega$ in the following manner:

$$\omega^* = \omega + \omega_R^{(x/y)} \frac{P}{1+\beta^2}, \quad (12)$$

$$\alpha^* = \frac{\alpha\omega + \beta\omega_R^{(x/y)}\frac{P}{1+\beta^2}}{\omega + \omega_R^{(x/y)}\frac{P}{1+\beta^2}}. \quad (13)$$

We note that the imaginary components of $\omega^*$ and $\omega$ are the same, as $\omega_R^{(x/y)}$ is entirely real. The stability of a phase was determined by the imaginary component of its eigenvalue, therefore any effect that RSOC may have on shifting the borders between a stable and an unstable region must be seen from $\alpha^*$. As $\omega$ is in general complex, so is $\alpha^*$ ($\alpha$, however, is real). It is also noted that the product $\alpha^*\omega^* = \alpha\omega + \omega_R^{(x/y)}P(1+\beta^2)^{-1}$ only has a shift in its real value. As $\omega$ and $\omega^*$ are complex and have the same imaginary component, the imaginary component of $\alpha^*$ must be non-zero if $\text{Re}(\alpha^*) \neq \alpha$. This imaginary component of $\alpha^*$ ensures that the imaginary component of $\alpha\omega$ is invariant under the transformation to $\alpha^*\omega^*$, but is otherwise of little interest as $\alpha$ only appears in the product $\alpha\omega$. The effect of RSOC can then be found from the real part of $\alpha^*$, which is found to be

$$\text{Re}(\alpha^*) = \frac{\left(\alpha\text{Re}(\omega) + \beta\omega_R^{(x/y)}\frac{P}{1+\beta^2}\right)\left(\text{Re}(\omega) + \omega_R^{(x/y)}\frac{P}{1+\beta^2}\right) + \alpha\Im(\omega)^2}{\left(\text{Re}(\omega) + \omega_R^{(x/y)}\frac{P}{1+\beta^2}\right)^2 + \Im(\omega)^2}. \quad (14)$$

There is no restriction on the sign of $\omega$ or $\omega_R^{(x/y)}$ ($\omega_R^{(y)}$ can for example be negative by switching the direction of the current $j_y$), and $\text{Re}(\alpha^*)$ can as a consequence also take on negative values. The effect of RSOC can then be seen as a modification of the Gilbert damping parameter $\alpha$, even to the extent where it takes on negative values and becomes an *anti-damping term* in the LLGS equation. We note that the modification $\alpha^*$ depends on the eigenvalues $\omega$, which are complex functions of the system parameters. As the ansatz is a small perturbation from a collinear state, the modified value $\alpha^*$ is only valid in this state, and will change again in a manner that cannot be described by this framework if the perturbation is unstable and we move away from the collinear state. One special case that should be noted is when $\alpha = \beta$, for which $\text{Re}(\alpha^*) = \alpha$. This predicts that when $\alpha = \beta$, RSOC has no impact on whether a collinear state is stable or not. When $\alpha \neq \beta$, $\text{Re}(\alpha^*)$ has a maximum and a minimum when $\Im(\omega) \neq 0$ and a singularity when $\Im(\omega) = 0$. These extremal points are found from

$$\frac{\partial}{\partial \omega_R}\text{Re}(\alpha^*) = -(\alpha - \beta) \times \frac{\Im(\omega)^2\left(\text{Re}(\omega) + 2\omega_R\frac{P}{1+\beta^2}\right) + \text{Re}(\omega)\left(\text{Re}(\omega) + \omega_R\frac{P}{1+\beta^2}\right)^2}{\left(\Im(\omega)^2 + \left(\text{Re}(\omega) + \omega_R\frac{P}{1+\beta^2}\right)^2\right)^2} = 0 \quad (15)$$

and are

$$\omega_R = -\left(\frac{1+\beta^2}{P}\right)\frac{\Im m(\omega)^2 \pm \Im m(\omega)\sqrt{\Im m(\omega)^2 + \text{Re}(\omega)^2} + \text{Re}(\omega)^2}{\text{Re}(\omega)}. \tag{16}$$

Which value is the maximum and minimum is determined by the sign of Re($\omega$), and whether $\beta > \alpha$ or $\beta < \alpha$. This means that if for our set of parameters we have Re($\alpha^*$) < $\alpha$ for $\alpha > \beta$, we will have Re($\alpha^*$) > $\alpha$ for the same set of parameters but with $\alpha < \beta$. Optimally we would like to have Re($\alpha^*$) < $\alpha$ to increase the size of the STO phase, and we would also like to be able to have this for both $\alpha < \beta$ and $\alpha > \beta$. This can be controlled by the direction of the current responsible for the SOC, as we would then switch the minimum and maximum in Re($\alpha^*$). Switching the sign of $\omega_R$ will have a similar effect to switching between $\alpha < \beta$ and $\alpha > \beta$.

In the bulk geometry switching the sign of $\omega_R$ can easily be done by switching the direction of the current in the $y$-direction. In the thin film geometry this may not have the desired effect, as the current controlling the strength of RSOC is also intertwined with the strength of the STT and by switching the direction of $j_x$ we will also change the value of $\omega$. The phase diagrams obtained numerically in the next section, determining when the STO phase occurs, are consistent with these analytical considerations of the role played by RSOC.

## Results

Based on the results presented so far, we will now calculate the phase diagrams to determine when the different collinear states are stable. This is done by solving (6) for all combinations of $\lambda_1$ and $\lambda_2$, and checking the sign of the imaginary component of $\omega$. In addition, to be able to separate the STO and canted phase from one another, we solve the LLGS equation numerically and analyze the results. We classify the system to be in an STO state if the variance of the latter part of the solution ($\mathbf{m}_0 \cdot \mathbf{m}_1$ and $\mathbf{m}_1 \cdot \mathbf{m}_2$) is above some minimal value (set to be $10^{-6}$ per time unit, defined as $\omega_0^{-1}$), and that the variance does not decrease faster than a cutoff factor (set to be 0.9) at an end interval of the solution. The precise value of these parameter values has little influence on the size of the STO phase as long as their magnitude is chosen reasonably (*i.e.* cutoff factors between 0 and 1 that are not too close to one of the end points, and a bound on the oscillations that does not discard oscillations of significant amplitude), as we will briefly discuss later. For an easy comparison to previous results, we use the same set of parameters as Zhou *et al.*[18] for our simulations, but additionally include the effect of spin–orbit torques. These parameters are $\alpha = 0.01$, $P = 0.5$, $K = 8 \cdot 10^4$ J/m$^3$, $d = 3$ nm, $J \sim 1$ mJ/m$^2$, $j_x \sim 10^8$ A/cm$^2$, $M_s = 127$ kA/m. We will also perform fast Fourier transforms of the numerical solution along given lines in the phase diagram, to analyze the effects of RSOC on the frequency spectrum.

**Bulk geometry.** A combination of the analytical and numerical calculations yield the results shown in Fig. 2 for the bulk geometry. We have here used a current density $j_y = 10^9$ A/cm$^2$ that is responsible for the SOC. The strength of the Rashba-parameter $\alpha_R$ is chosen large enough to have considerable impact, but kept at a realistic order of magnitude. As an example, $\alpha_R$ has been found to be $\sim 3.7 \cdot 10^{-10}$ eV·m at the surface of Bi/Ag alloy[23]. The value $\alpha_R = 9.26 \cdot 10^{-10}$ eV·m exceeds this value and thus corresponds to a rather large value of the Rashba-parameter. For the bulk geometry this is of no concern, as we can achieve the same results for lower values of $\alpha_R$ by increasing the current density $j_y$, but this is not possible in the thin film geometry where spin-transfer torques and RSOC can not be separated in the same manner.

To benchmark our numerical results, we reproduce the phase diagram found by Zhou *et al.* in the absence of spin–orbit interactions in Fig. 2a. When Rashba spin–orbit coupling is present, it is seen from Fig. 2b that the STO phase becomes larger. In addition, the STO phase also extends into the antiferromagnetic regime ($J < 0$), and even occurs in the absence of any RKKY interaction between $\mathbf{m}_1$ and $\mathbf{m}_2$ ($J = 0$). In the article by Zhou *et al.*, the border between the unstable region (STO and canted phases) and the ↑↓ state was found to be approximately

$$\omega_J = \sqrt{4\omega_0^2 + (\omega_j^{(x)})^2} - 2\omega_0 + \alpha\frac{\omega_0^2}{\omega_j^{(x)}}. \tag{17}$$

The effect of the term proportional to $\alpha$ can be seen by the slight increase of the border between the unstable region and the ↑↓ state as $j_x \to 0^+$ in Fig. 2a. In Fig. 2b, however, the border decreases as $j_x \to 0^+$. This is in agreement with what we have discussed earlier, namely that the effect of RSOC in the present system is equivalent to a modification of the Gilbert damping, even to the extent where it takes on negative values. This seems to be the case in this phase diagram where Re($\alpha^*$) becomes minimal when $\beta < \alpha$. We also predicted above that one could move from Re($\alpha^*$) < $\alpha$ to Re($\alpha^*$) > $\alpha$ by letting $\beta > \alpha$, assuming Re($\alpha^*$) < $\alpha$ for $\beta < \alpha$. This is illustrated in Fig. 2c,d, where the border between the unstable region and the ↑↓ state is lifted with respect to the border in Fig. 2a. The size of the STO phase also decreases in this maximal region of Re($\alpha^*$) (Re($\alpha^*$) > $\alpha$), which is not an unexpected consequence from an increase in the Gilbert damping. This does not mean that cases where $\beta > \alpha$ are of no interest, however. As noted, we are able to move back to a minimal region of Re($\alpha^*$) in the bulk geometry by switching the direction of the current applied in the $y$-direction.

To illustrate further the effects that the spin–orbit torques resulting from RSOC may have on the phase diagram, we also find the phase diagram in the $J/j_y$-plane for zero applied current in the $x$-direction (no effects from STT). This is illustrated in Fig. 3. As we can see, the SOT alone is not able to generate an STO phase. For the size of the STO phase to increase the STT and SOT must work in unison. However, we predict that the SOT will affect the stability of the different collinear states, where ↓↓ is the dominant stable state for $j_y \ll 0$ and ↑↑ the dominant state for $j_y \gg 0$. This is even valid in the antiferromagnetic regime, assuming a sufficiently small RKKY coupling $J$. We see that for small $j_y$ the antiferromagnetic collinear states are still the stable ones for $J < 0$, and this phase seems to be increasing as the antiferromagnetic coupling strength increases. For small $j_y$ and $J$ all collinear states are stable. The transition from the ↓↓ state to the ↑↑ state by changing the sign of $j_y$ in the absence of a current in the $x$-direction is a nice check to see whether the separation of STT from SOT is feasible. For a small RKKY coupling between F$_1$ and F$_2$ and large current densities, the STT case with a current in the $x$-direction should transition from having ↑↑ as the stable state to ↓↑ as $j_x \ll 0 \rightarrow j_x \gg 0$. The SOT case, however, with a current in the $y$-direction should transition from ↓↓ to ↑↑ as $j_y \ll 0 \rightarrow j_y \gg 0$.

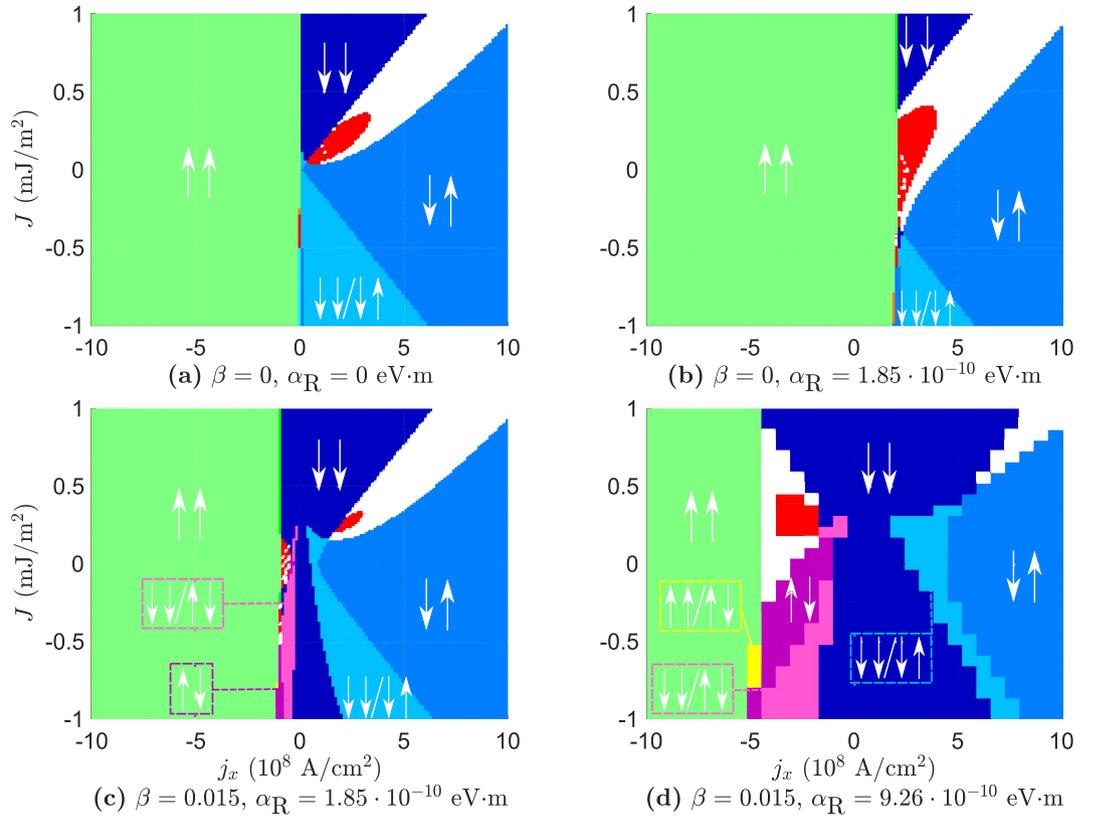

**Figure 2. Phase diagrams for the bulk geometry.** For all phase diagrams the values $\alpha = 0.01$, $P = 0.5$, $j_y = 10^9$ A/cm$^2$, $K = 8 \cdot 10^4$ J/m$^3$, $d = 3$ nm are assumed. The red regions indicate an STO phase, while the white regions are canted states.

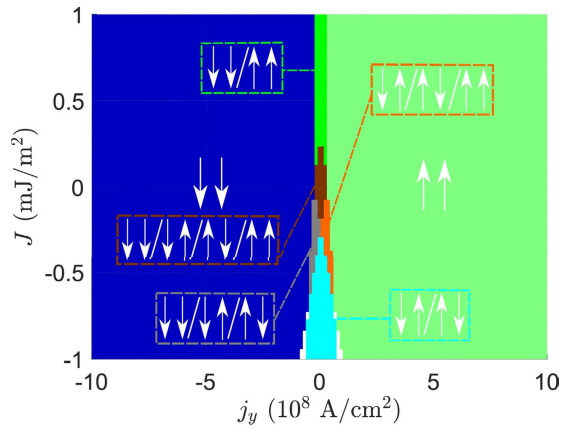

**Figure 3. Phase diagram for the bulk geometry as a function of $j_y$ ($j_x = 0$ A/cm$^2$, $\alpha_R = 9.26 \cdot 10^{-10}$ eV·m).** The white regions are canted states, and there is no STO phase.

Moving on to the frequency spectrum of the STO phase in the presence of RSOC, we consider fast Fourier transforms of the quantity $\mathbf{m}_0 \cdot \mathbf{m}_1$ along the $J = 0.25$ mJ/m$^2$ line in the phase diagrams as a function of $j_x$. These results are presented in Fig. 4. The fast Fourier transforms of $\mathbf{m}_1 \cdot \mathbf{m}_2$ are not presented as they show the same frequency spectrum as $\mathbf{m}_0 \cdot \mathbf{m}_1$, but with different amplitudes. The system with RSOC has fewer frequencies, as can be seen by comparing Fig. 4a,b. In addition, the oscillations in the system with RSOC can have a higher amplitude than the system without RSOC, as illustrated in Fig. 4b,d. It is also seen that the presence of RSOC allows us to achieve slightly different frequency outputs than a system without RSOC. This will increase the tunability of the spin torque oscillator, as we can also adjust the strength of the RSOC via $j_y$, in addition to the modulation of the frequency via $j_x$ and the anisotropy strength $K$ as done in ref. 18.

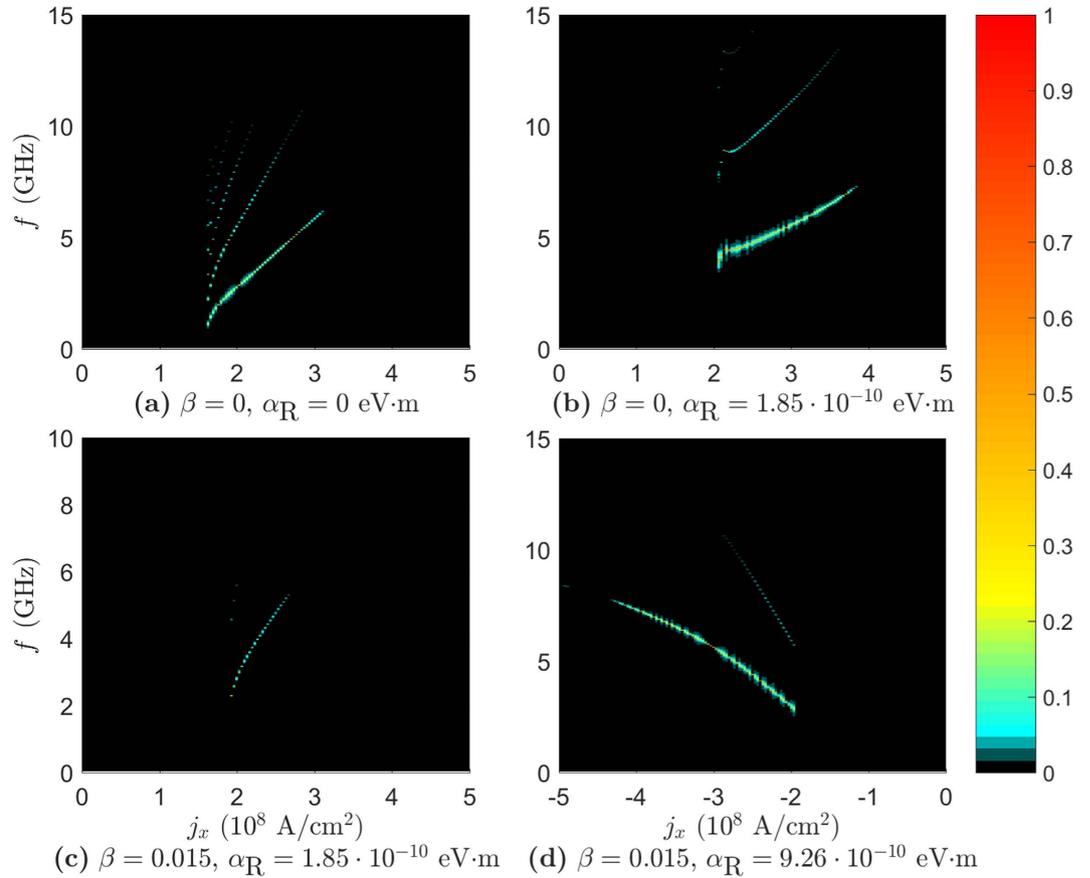

**Figure 4. The one sided spectra of the Fourier transform of $m_0 \cdot m_1$ in the bulk geometry.** The Fourier transform was taken along the $J = 0.25$ mJ/m$^2$ line in the phase diagrams.

**Thin film geometry.** In the thin film geometry, both the spin-transfer torques and spin–orbit coupling are controlled by the same current $j_x$. This gives us a different behavior than the one in the bulk geometry, as seen in Fig. 5. The reason for this is that the phase diagrams in the bulk geometry are influenced uniformly by the spin–orbit torques due to the current $j_y$, while in the thin film geometry both STT and SOT are controlled by the same current $j_x$. This leads to a non-uniform influence from SOT on the phase diagrams, and for large values of $\alpha_R$ a significant change from a system without RSOC to one with RSOC can be observed, as in Fig. 5a. As $\beta \rightarrow \alpha$ the effects of RSOC become less apparent (vanishing completely at $\beta = \alpha$), as in Fig. 5b. For low values of $\alpha_R$ the behavior of the thin film geometry is quite similar to a system without RSOC, but as we increase $\alpha_R$ interesting effects appear. When we let $\alpha_R = 9.26 \cdot 10^{-10}$ eV·m and $\beta = 0$, we get the phase diagram shown in Fig. 5a. In this phase diagram the STO phase is symmetric in $j_x$, and the STO phase is located in the antiferromagnetic regime ($J < 0$). When we let $\beta = \alpha/2 = 0.005$ the phase diagram becomes similar to the case without any RSOC again, but there is still a significant increase in the STO phase which extends into the antiferromagnetic regime. The decrease in the border between the unstable region and the ↑↓ state as $j_x \rightarrow 0^+$ is also not existent in the thin film geometry, as the effects of RSOC are proportional to $j_x$. The significant difference between Fig. 5a,b is due to that for low values of $\beta$, the phase diagram shifts in the sign of $j_x$ for very large values of $\alpha_R$, so that the phase diagram resembles more the phase diagram without the presence of RSOC but with $j_x \rightarrow -j_x$. For our choice of system parameters, the value $\alpha_R = 9.26 \cdot 10^{-10}$ eV·m marks the mid-point between the transition of having the STO phase for positive $j_x$ to having it for negative $j_x$, and thus gives the largest STO phase.

The STO phase in Figs 2 and 5 is only registered where there are no stable collinear states according to the analytical calculations, as we only want to classify a state as being STO if there is not a possibility of the system stabilizing in a collinear state. The stability of the collinear states are determined analytically, while the STO phase is found numerically. If we include all the points the numerical calculations registered as being an STO phase, even at points where the analytical calculations show one or more collinear states to be stable, the phase diagrams become like the ones shown in Fig. 6.

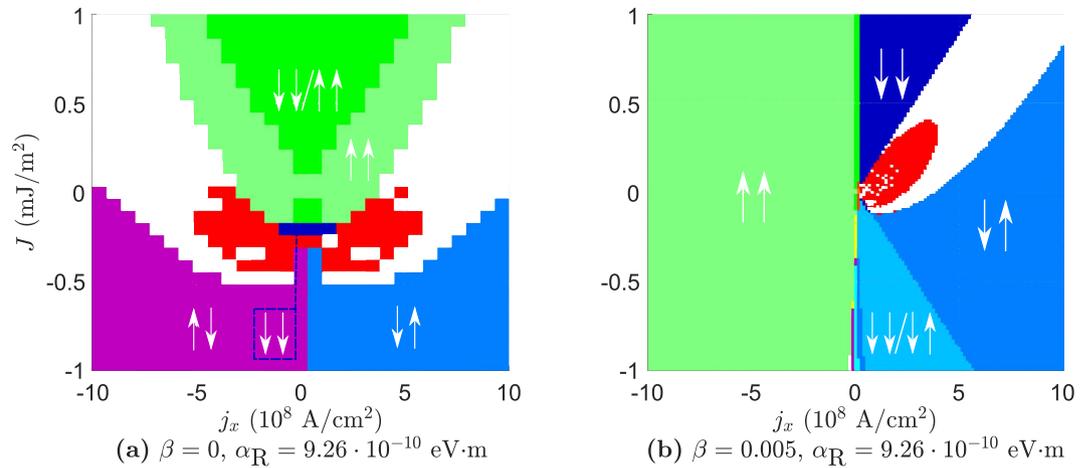

**Figure 5. Phase diagrams for the thin film geometry.** For all phase diagrams the values $\alpha = 0.01$, $P = 0.5$, $K = 8 \cdot 10^4$ J/m$^3$, $d = 3$ nm are assumed. The red regions indicate an STO phase, while the white regions are canted states.

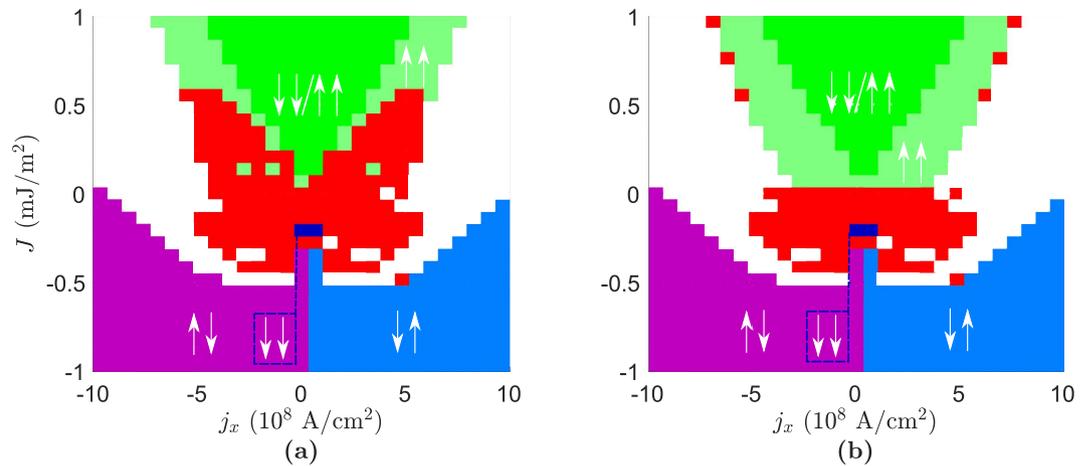

**Figure 6. The stability of the ↑↑ state in the thin film geometry for $\beta = 0$, $\alpha_R = 9.26 \cdot 10^{-10}$ eV·m is found to be very dependent on the initial conditions.** In (**a**) the initial state is chosen to be close to ↓↓ for $J > 0$, while in (**b**) the initial state is chosen to be close to ↑↑ in the same region. In both cases the initial state is chosen to be close to ↓↑ for $J < 0$.

We see that the STO phase also extends into the region where ↑↑ is a stable state when the initial state is not close to this state. It is therefore also possible to get an STO phase in the ferromagnetic regime for this case, depending on the initial state of $\mathbf{m}_1$ and $\mathbf{m}_2$. The reason why the initial state does not relax to the ↑↑ state even if this is a stable solution can be seen from the size of the imaginary component of $\omega$ in this region. While the imaginary component of $\omega$ belonging to ↑↑ in this region is negative, its magnitude is only $10^{-2}$. In comparison the other collinear states have an imaginary component ranging up to a magnitude of $10^1$. If $\mathbf{m}_1$ and $\mathbf{m}_2$ are not close to being in the ↑↑ state initially, it is therefore not given that they will relax to this state, and can end up in an oscillating or canted state.

This sensitivity issue concerning the initial state of $\mathbf{m}_1$ and $\mathbf{m}_2$ was only of considerable significance for the ↑↑ state in the thin-film geometry with $\beta = 0$, $\alpha_R = 9.26 \cdot 10^{-10}$ eV·m. There was also some sensitivity for other phase diagrams in the limit $j_x \to 0$. Regarding the cutoff factors utilized for classifying the numerical solution, there was very little sensitivity with regard to the choice of parameters. The transitional region between an STO state and a canted or collinear state is found to happen on a typical scale of $10^6$ A/cm$^2$ or $10^{-3}$ mJ/m$^2$, both of which being very small compared to the magnitude of the parameters considered. As the cutoff factors would primarily have an effect in this transitional region, a change in these parameters would not lead to a significant change in the size of the STO phase. It can also be seen in some of the phase diagrams that there are found to be some canted states in the middle of the STO region, or a single STO state separated from the main STO region. This is discovered to be due to the time interval the system is solved over, as for some parameters it takes longer time for the system to obtain stable oscillations than for others. The time scale that the system was solved over was constrained to some degree, due to the run-time of the solver. It is also observed that the initial state of $\mathbf{m}_1$ and $\mathbf{m}_2$ has some impact on how fast these oscillations become stable. When solving the system over a longer time scale the STO regions should become more or less continuous.

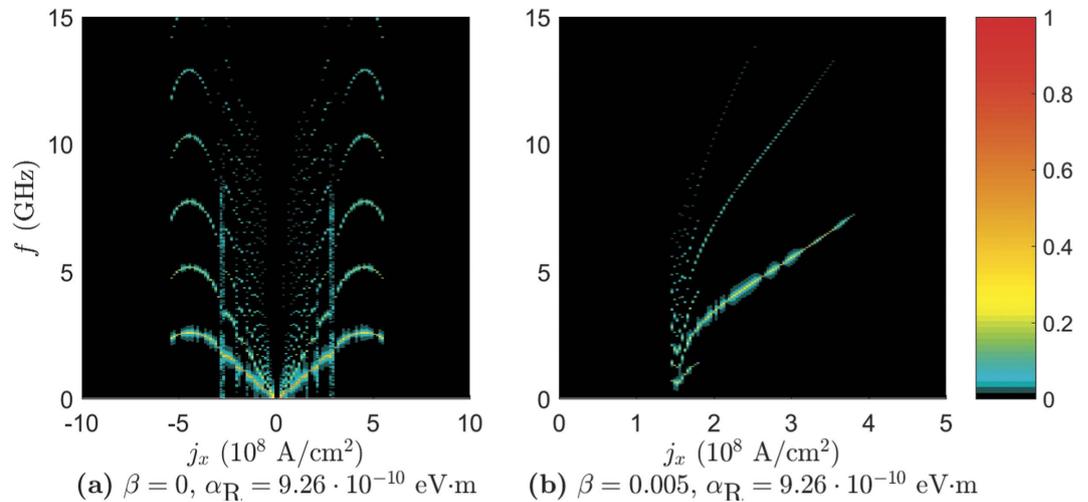

**Figure 7. The one sided spectra of the Fourier transform of $m_0 \cdot m_1$ in the thin film geometry.** The Fourier transform was taken along the $J = -0.2\,\text{mJ/m}^2$ line in (**a**) and along $J = 0.25\,\text{mJ/m}^2$ in (**b**).

When considering the frequency spectrum of the STO phase in the thin film geometry in Fig. 7 we see that the spectrum of the $\alpha_R = 9.26 \cdot 10^{-10}\,\text{eV} \cdot \text{m}$, $\beta = 0$ case shows the same symmetry in $j_x$ as the phase diagram. Moreover, the antiferromagnetic STO phase has a larger set of frequencies than what has been found in the ferromagnetic STO phase and is also, unlike the other plots, symmetric in $j_x$. When increasing $\beta$ to 0.005 and considering the frequency spectrum in the ferromagnetic regime, the result is more similar to the bulk geometry case, as seen in Fig. 7b. We have fewer and more slowly varying frequencies as a function of $j_x$, but the amplitude of the oscillations are higher.

## Conclusion

We have shown how Rashba spin–orbit coupling can be used to substantially increase the size of the spin torque oscillator phase in both ferromagnetically and antiferromagnetically coupled compensated magnetic moments in a bilayer system. This presumably allows for a better tunability of the frequency output of the oscillator, as the spin–orbit coupling torques can be controlled electrically.

### Acknowledgements

J.L. was supported by the Research Council of Norway, Grants Nos 216700 and 240806 and the "Outstanding Academic Fellows" programme at NTNU.

### Author Contributions

Ø.J. did the analytical and numerical calculations with support from J.L. Both authors contributed to the discussion of the results and the writing of the manuscript.

### Additional Information

**Competing financial interests:** The authors declare no competing financial interests.

**How to cite this article**: Johansen, Ø. and Linder, J. Current driven spin–orbit torque oscillator: ferromagnetic and antiferromagnetic coupling. *Sci. Rep.* **6,** 33845; doi: 10.1038/srep33845 (2016).